\documentclass[conference]{IEEEtran}
\IEEEoverridecommandlockouts

\usepackage{cite}
\usepackage{amsmath,amssymb,amsfonts}
\usepackage{algorithmic}
\usepackage{graphicx}
\usepackage{textcomp}
\usepackage{xcolor}
\usepackage{hyperref}
\usepackage{siunitx}
\usepackage{hyperref}
\usepackage{tikz}

\usepackage{xcolor}

\usepackage{lipsum}
\usepackage[pscoord]{eso-pic}
\newcommand{\placetextbox}[3]{
  \setbox0=\hbox{#3}
  \AddToShipoutPictureFG*{
    \put(\LenToUnit{#1\paperwidth},\LenToUnit{#2\paperheight}){\vtop{{\null}\makebox[0pt][c]{#3}}}%
  }
}

\usepackage{siunitx}

\def\BibTeX{{\rm B\kern-.05em{\sc i\kern-.025em b}\kern-.08em
    T\kern-.1667em\lower.7ex\hbox{E}\kern-.125emX}}
\begin{document}
\placetextbox{0.5}{1}{This is the author's version of an article that has been published.}
\placetextbox{0.5}{0.985}{Changes were made to this version by the publisher prior to publication.}
\placetextbox{0.5}{0.97}{The final version of record is available at \href{https://doi.org/10.1109/WFCS63373.2025.11077640}{https://doi.org/10.1109/WFCS63373.2025.11077640}}%
\placetextbox{0.5}{0.05}{Copyright (c) 2025 IEEE. Personal use is permitted.}
\placetextbox{0.5}{0.035}{For any other purposes, permission must be obtained from the IEEE by emailing pubs-permissions@ieee.org.}%

\title{Analysis of the operation of a TSN switch and other devices using executable QR codes\vspace{-0.3cm}\\
\thanks{This work has been partially funded by CNR under the project ``Executable QR codes (EQR) - GORU IEIIT'' (DIT.AD001.212).}
}

\author{
    \IEEEauthorblockN{
    Stefano Scanzio,
    Pietro Chiavassa,
    Gianluca Cena
    }  
    \IEEEauthorblockA{National Research Council of Italy (CNR--IEIIT), Italy. }    
    Email:  \{stefano.scanzio, pietrochiavassa, gianluca.cena\}@cnr.it
    \vspace{-0.3cm}
}

\maketitle

\begin{abstract}
Executable QR codes, also known as sQRy, are a technology aimed at inserting executable programs in a QR code.

Through a concrete example, in this paper, we demonstrate their usage in the context of industrial networks in order to assess the operation of a TSN switch by analyzing its status LEDs even in the absence of an internet connection.

The entire generation chain that is used to create the sQRy, as well as the corresponding execution chain that, starting from the sQRy, runs it on a mobile device, has been detailed through examples.
\end{abstract}

\section{Introduction and Software Architecture}
Executable QR codes (eQR code) or sQRy (\url{https://www.sqry.org}) were invented in 2022 \cite{9921530}, and allow 
an executable program
to be embedded in a QR code. 
This permits 
sQRy execution also in the absence of an Internet connection, a condition that is not unlikely in several kinds of industrial installations
(including those that contain local networks and the related equipment), 
e.g., in remote areas (such as petrochemical plants) or in basement floors of buildings. 
The main challenge in the generation of a sQRy is to make the program compact enough to be included in the storage capacity of a QR code (\SI{2953}{bytes} at most). 
For this reason, a specific language termed QRtree, which is called dialect in sQRy terminology, was proposed with the target of having a compact solution aimed at coding decision trees. 
In the context of QRtree, a decision tree is a program that interacts with the user through a sequence of questions and, depending on the answers, it can provide suitable feedback to the user. 
Another dialect, we termed QRind, 
has been explicitly conceived for industrial applications, but at the moment of writing it is not fully implemented yet \cite{10710739}.

The QRtree dialect was presented in \cite{10492995}, in the context of mountain trails, and fully detailed 
in specification documents \cite{QRscript-spec, QRtree-spec}.
The software components behind the sQRy technology can be subdivided into two steps: the \textit{generation chain} that is used to embed a program in the QR code (i.e., the sQRy), and the \textit{execution chain} that starts from the acquisition of the sQRy through a camera and continues with the execution, by means of emulation, of the embedded program.

\begin{figure}[b]
\vspace{-0.5cm}
\begin{center}
\includegraphics[width=0.4\linewidth]{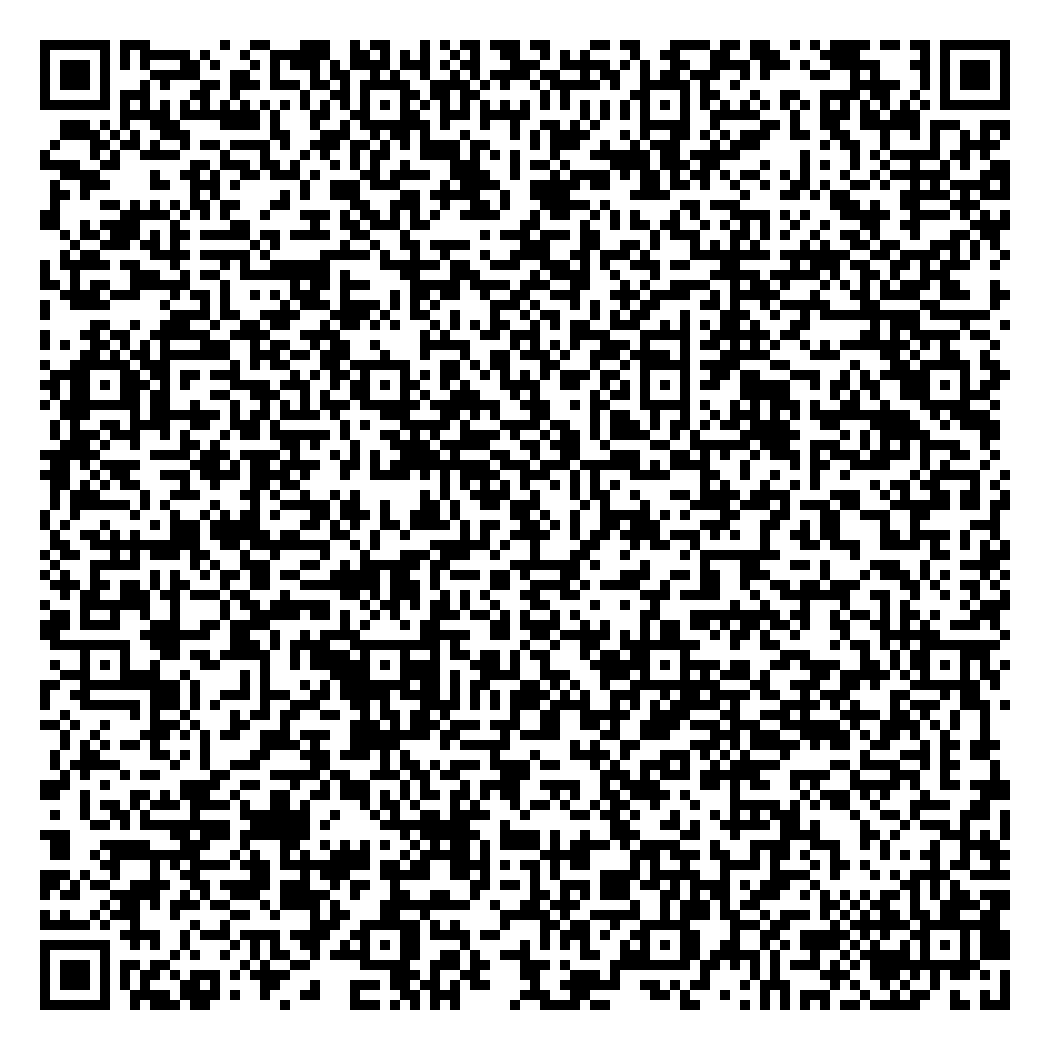}
\end{center}\
\vspace{-0.9cm}
\caption{sQRy of the complete demo program.}
\label{fig:sQRy}
\end{figure}

In the generation chain, the source code (written using a high-level programming language) is translated into an intermediate representation, then the intermediate representation is converted into a binary format and, finally, the stream of bytes is embedded into a conventional QR code obtaining the sQRy (see Fig.~\ref{fig:sQRy} for an example of sQRy).

The execution chain starts from the sQRy printed on a support
(label, poster, tag, etc.). 
A specialized app detects and extracts the content from the QR code, obtaining its binary representation. This binary data is then converted back into an intermediate format. Finally, the program is executed by a virtual machine on the user's device, which supports user interaction (e.g., by using a touchscreen).

\section{Demonstration}
A specific Time-Sensitive Networking (TSN) switch, model NZ2MHG-TSNT8F2 \cite{mitsubishi2025tsnswitch}, 
produced by Mitsubishi Electric, has been selected as the reference device for this demo. 
However, many other types of industrial network equipment (and even unconnected devices) could serve as meaningful examples.
In the proposed setup, a sQRy based on the QRtree dialect is stuck on the switch enclosure to provide the user information about the switch's status based on its LEDs. 
Similar programs can be used to provide descriptions about the configuration parameters of the specific switch (e.g., how to access it, what is the IP address assigned to the switch by the factory network administrator), to assist in the resolution of operational issues or to provide an advanced smart user guide.

The switch we analyzed has $5$ LEDs on its frontal panel (\texttt{RUN}, \texttt{ERR}, \texttt{PW1}, \texttt{PW2}, \texttt{SYNC}), one LED for each one of its $8$ Ethernet ports, and one LED for each one of its optical fiber ports. 
For space reasons, only the LEDs \texttt{RUN} and \texttt{ERR} on the frontal panel will be analyzed below. 
The complete example about all LEDs in the frontal panel is available on \url{https://github.com/eQR-code/QRtree/tree/main/examples} (see files starting with \texttt{ex03}).

\begin{figure}[t]
\fontsize{7}{8}\selectfont
\linespread{0.85}
\begin{verbatim}
input "What led?"
if "RUN LED":
    print "Operating status of the switch"
    input "What color?"
    if "Green":
        print "Normal operation" exit
    else if "Flashing Green":
        input "At what speed?"
        if "500 ms interval":
            print "Reset button pressed" exit
        else if "250 ms interval":
            print "Normally operating with
            USB drive connected"
            exit
    else if "Flashing Red":
        print "Initializing" exit
    else if "Off":
        print "Power-off" exit

else if "ERR LED":
    print "Error status"
    input "What color?"
    if "On Red":
        print "Initial error occurred/USB
        flash drive failed" exit
    if "Off Red":
        print "No error" exit
\end{verbatim}
\vspace{-0.2cm}
\caption{High-level representation of the code included in the sQRy.}
\vspace{-0.1cm}
\label{fig:high_level}
\end{figure}

The high-level representation of the code included in the sQRy is sketched in Fig.~\ref{fig:high_level}.
In detail, the program asks the user to identify a specific LED. 
Then, it prints a brief description of the LED function(s) and asks the user for information about the LED color. 
Finally, it displays a clear textual explanation about the current device's status (as conveyed by the LED).
\begin{figure}[t]
\fontsize{7}{8}\selectfont
\linespread{0.85}
\begin{verbatim}
(0) input "What led?"
(1) if "RUN LED" (4)
(2) if "ERR LED" (23)
(3) goto (31)
(4) print "Operating status of the switch"
(5) input "What color?"
(6) if "Green" (11)
(7) if "Flashing Green" (13)
(8) if "Flashing Red" (20)
(9) if "Off" (21)
(10) goto (22)
(11) printex "Normal operation"
(12) goto (22)
(13) input "At what speed?"
(14) if "500 ms interval" (17)
(15) if "250 ms interval" (19)
(16) goto (20)
(17) printex "Reset button pressed"
(18) goto (20)
(19) printex "Normally operating with USB drive connected"
(20) printex "Initializing"
(21) printex "Power-off"
(22) goto (31)
(23) print "Error status"
(24) input "What color?"
(25) if "On Red" (27)
(26) goto (28)
(27) printex "Initial error occurred/USB flash drive failed"
(28) if "Off Red" (30)
(29) goto (31)
(30) printex "No error"
\end{verbatim}
\vspace{-0.3cm}
\caption{Part of the intermediate representation.}
\label{fig:intermediate}
\vspace{-0.3cm}
\end{figure}
This program is translated into the intermediate representation of Fig.~\ref{fig:intermediate}\footnote{A prototype open-source software implementation of the generation and execution chain is available: \url{https://github.com/eQR-code/QRtree}.}, and then it is inserted in the QR code shown in Fig.~\ref{fig:sQRy}. The intermediate code is composed of only $7$ simple instructions. 
The produced sQRy has a payload of \SI{2720}{bits}, which corresponds to an occupation of $11.5\%$ of its maximum allowed capacity.

\begin{figure}[t]
\begin{center}
\includegraphics[width=0.85\linewidth]{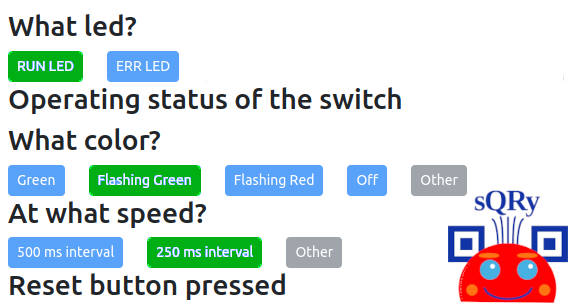}
\end{center}
\vspace{-0.4cm}
\caption{Example of execution of the demo program.}
\label{fig:interface}
\vspace{-0.5cm}
\end{figure}

An example of the graphical interface presented to the user when executing the sQRy is reported in Fig.~\ref{fig:interface}, which is an HTML page based on \textit{JavaScript} that is automatically generated by the execution chain after scanning the sQRy in Fig.~\ref{fig:sQRy}. 
This interface can be used in practically any device equipped with a web browser, including a smartphone application. 
Concerning the example in Fig.~\ref{fig:interface}, we assume that the user has selected the \texttt{RUN} LED, which is used to communicate the ``Operating status of the switch''. 
Since its color is ``Flashing Green'' and the flashing interval is \SI{250}{ms}, the program returns the message ``Reset button pressed".

\section{Conclusions}
A demonstration of the use of a sQRy (eQR code) is presented for easing the analysis of the status of a TSN switch based on the configuration of its status LEDs. 
This is a representative example that shows the effectiveness of this technology when applied in the context of configuration and troubleshooting of industrial communication equipment.

Although sQRy is a key enabler in many application contexts, it has some limitations. 
The amount of data that can be stored in a QR code limits the size of the program. 
As already described in the specification documents, this problem could be solved by using multiple QR codes that are scanned before program execution or at specific points in the decision tree. 
The latter option is part of future work.

An alternative solution to sQRy would be to have a dedicated app embedding support for a set of devices. This can be done, e.g., by inserting all the relevant functions inside the app, and an identifier in a QR code pointing to the function to be invoked.
However, this requires the app to be maintained, which can be a burden for the administrator, and the producer of the sQRy and the app must coincide.
Moreover, the app may become unavailable after the device is discontinued.
Instead, with an open sQRy standard, it will always be possible to run the software.

\bibliographystyle{IEEEtran}
\bibliography{bibliography}

\begin{thebibliography}{1}
\providecommand{\url}[1]{#1}
\csname url@samestyle\endcsname
\providecommand{\newblock}{\relax}
\providecommand{\bibinfo}[2]{#2}
\providecommand{\BIBentrySTDinterwordspacing}{\spaceskip=0pt\relax}
\providecommand{\BIBentryALTinterwordstretchfactor}{4}
\providecommand{\BIBentryALTinterwordspacing}{\spaceskip=\fontdimen2\font plus
\BIBentryALTinterwordstretchfactor\fontdimen3\font minus
  \fontdimen4\font\relax}
\providecommand{\BIBforeignlanguage}[2]{{%
\expandafter\ifx\csname l@#1\endcsname\relax
\typeout{** WARNING: IEEEtran.bst: No hyphenation pattern has been}%
\typeout{** loaded for the language `#1'. Using the pattern for}%
\typeout{** the default language instead.}%
\else
\language=\csname l@#1\endcsname
\fi
#2}}
\providecommand{\BIBdecl}{\relax}
\BIBdecl

\bibitem{9921530}
S.~Scanzio, G.~Cena, and A.~Valenzano, ``{QRscript: Embedding a Programming
  Language in QR codes to support Decision and Management},'' in \emph{2022
  IEEE 27th International Conference on Emerging Technologies and Factory
  Automation (ETFA)}, 2022, pp. 1--8.

\bibitem{10710739}
S.~Scanzio, F.~Velluto, M.~Rosani, L.~Wisniewski, and G.~Cena, ``{Executable QR
  codes with Machine Learning for Industrial Applications},'' in \emph{2024
  IEEE 29th International Conference on Emerging Technologies and Factory
  Automation (ETFA)}, 2024, pp. 1--4.

\bibitem{10492995}
S.~Scanzio, M.~Rosani, M.~Scamuzzi, and G.~Cena, ``{QR Codes: From a Survey of
  the State of the Art to Executable eQR Codes for the Internet of Things},''
  \emph{IEEE Internet of Things Journal}, vol.~11, no.~13, pp.
  23\,699--23\,710, 2024.

\bibitem{QRscript-spec}
\BIBentryALTinterwordspacing
------, ``{{QRscript specification}},'' \emph{arXiv}, pp. 1--13, Mar. 2024.
  [Online]. Available: \url{https://doi.org/10.48550/arXiv.2403.04708}
\BIBentrySTDinterwordspacing

\bibitem{QRtree-spec}
\BIBentryALTinterwordspacing
------, ``{{QRtree - Decision Tree dialect specification of QRscript}},''
  \emph{arXiv}, pp. 1--32, Mar. 2024. [Online]. Available:
  \url{https://doi.org/10.48550/arXiv.2403.04716}
\BIBentrySTDinterwordspacing

\bibitem{mitsubishi2025tsnswitch}
\BIBentryALTinterwordspacing
{Mitsubishi Electric}, ``{NZ2MHG-TSNT8F2 TSN switch},'' 2025, accessed: May
  5th, 2025. [Online]. Available:
  \url{https://it.mitsubishielectric.com/fa/it_en/products/cnt/plc/plcccl/ethernet/nz2mhg-tsnt8f2.html}
\BIBentrySTDinterwordspacing

\end{thebibliography}

\cleardoublepage

\end{document}